\documentclass[showpacs,preprint,amsmath,amssymb]{revtex4}
\usepackage{graphicx}

\def\ams#1#2#3{{ #3 }{\textit{Ann. Math. Stat.} } {\bf{#1}} #2}

\def\arpc#1#2#3{{ #3 }{\textit{Annu. Rev. Phys. Chem.} } {\bf{#1}} #2}

\def\jas#1#2#3{{ #3 }{\textit{J. Am. Chem. Soc.} } {\bf{#1}} #2}
\def\japs#1#2#3{{ #3 }{\textit{J. Appl. Polym. Sci.} } {\bf{#1}} #2}

\def\jcp#1#2#3{{ #3 }{\textit{J. Chem. Phys. }} {\bf{#1}} #2}

\def\jmsc#1#2#3{{ #3 }{\textit{J. Macromol. Sci: Chem. }} {\bf{#1}} #2}

\def\jpc#1#2#3{{ #3 }{\textit{J. Phys. Chem.} } {\bf{#1}} #2}

\def\jps#1#2#3{{ #3 }{\textit{J. Polym. Sci.} } {\bf{#1}} #2}
\def\jpsc#1#2#3{{ #3 }{\textit{J. Polym. Sci., C: Polym. Lett.} } {\bf{#1}} #2}
\def\jpsa#1#2#3{{ #3 }{\textit{J. Polym. Sci., A: Polym. Chem.} } {\bf{#1}} #2}

\def\macro#1#2#3{{ #3 }{\textit{Macromolecules} } {\bf{#1}} #2}
\def\mts#1#2#3{{ #3 }{\textit{Macromol. Theory Simul.} } {\bf{#1}} #2}
\def\macroc#1#2#3{{ #3 }{\textit{Makromolecular Chemie} } {\bf{#1}} #2}

\def\pnas#1#2#3{{ #3 }{\textit{Proc. Natl. Acad. Sci. USA} } {\bf{#1}} #2}

\def\jsm#1#2#3{{ #3 }{\textit{J. Stat. Mech.} } {\bf{#1}} #2}
\def\pb#1#2#3{{ #3 }{\textit{Phys. Biol.} } {\bf{#1}} #2}
\def\jcp#1#2#3{{ #3 }{\textit{J. Chem. Phys.} } {\bf{#1}} #2}

\begin{document}

\title{A general theory of kinetics and thermodynamics of steady-state copolymerization}

\author{Yao-Gen Shu$^{a,b}$}
\author{ Yong-Shun Song$^b$}
\author{Zhong-Cun Ou-Yang$^a$}
\author{Ming Li$^b$}\email{liming@ucas.ac.cn}
\affiliation {$^{a}$~State Key Laboratory of Theoretical Physics \\
Institute of Theoretical Physics, Chinese Academy of Sciences\\
$^{b}$~School of Physics, University of Chinese Academy of Sciences\\
No.19A Yuquan Road, Beijing 100049, P. R. China}
\date{\today}

\begin{abstract}
Kinetics of steady-state copolymerization has been investigated since 1940s. Irreversible terminal and penultimate models were successfully applied to a number of comonomer systems, but failed for systems where depropagation is significant. Although a general mathematical treatment of the terminal model with depropagation was established in 1980s, penultimate model and higher-order terminal models with depropagation have not been systematically studied, since depropagation leads to hierarchically-coupled and unclosed kinetic equations which are hard to be solved analytically. In this work, we propose a truncation method to solve the steady-state kinetic equations of any-order terminal models with depropagation in an unified way, by reducing them into closed steady-state equations which give the exact solution of the original kinetic equations. Based on the steady-state equations, we also derive a general thermodynamic equality in which the Shannon entropy of the copolymer sequence is explicitly introduced as part of the free energy dissipation of the whole copolymerization system.
\end{abstract}
\maketitle


\section{Introduction} \label{Introduction}
Understanding the kinetics of copolymerization and thus controlling the copolymer sequence statistics (e.g., copolymer composition, sequence distribution) are the key subjects in the study of copolymer, since the sequence statistics significantly affect the chemical and physical properties of the copolymers \cite{handbook2004}.
Therefore, it became an important issue to theoretically model the copolymerization kinetics and estimate the rate constants of all the involved polymerization reactions. This has drawn a lot of attention both experimentally and theoretically since 1940s. In order to study the kinetics, the experiments are usually conducted at low conversion conditions to maintain the copolymerization process at steady state (i.e., the monomer concentrations in the environment are almost unchanged during the process), which actually much simplifies the modeling and analysis of experiment data. Based on the steady-state assumption, different theoretical models have been suggested for different systems. Early works assumed the so-called terminal effects, i.e., the last monomer unit at the growing end of the copolymer influences the chain growth and thus the copolymer composition. Several terminal models were developed in 1940s and successfully applied to experiments \cite{wall1941,mayo1944,alfrey1944,AG1944}.
Besides the assumptions of terminal effect, these early models also assumed that the copolymerization reactions are irreversible, which ensures the corresponding kinetic equations to be solved analytically. These two assumptions were shown insufficient to explain later experimental results, which leads to the development of two other categories of models.

The first category was proposed to account for the so-called penultimate effect, i.e., the next-to-last (penultimate) monomer unit at the growing end can have substantial influence on the copolymerization kinetics (e.g., \cite{fukuda1985,davis1989}).
The original penultimate model was suggested by Merz et al. \cite{merz1946}, and then was revised and developed (for a review, see Ref.\cite{coote2000}). Besides the terminal (also called as first-order terminal in this article) and penultimate (the second-order terminal) effects, higher-order terminal effects are also possible (e.g., antepenultimate effect \cite{lowry1960}). But such cases have not been systematically investigated.

The second category was proposed to account for depropagation effect which brings substantial mathematical difficulty to the studies of copolymerization kinetics. Depropagation was noticed very early in 1960s. It originated from the thermodynamic argument, i.e., all the reactions pathways are essentially reversible and depropagation may become significant at some elevated temperature. A few copolymerization systems do exhibit depropagation which shows substantial impacts on the copolymerization kinetics and copolymer composition (e.g., \cite{driscoll1967,guillot1997,penlidis2005}). Such temperature effects can only be described by reversible models. However, depropagation always leads to hierarchically-coupled and unclosed kinetic equations which are hard to be solved analytically (as will be clear in later sections). Because of this mathematical difficulty, it was until 1987 that the first systematic treatment of first-order terminal models with depropagation was given by Kruger et.al. \cite{kruger1987}. Kruger's approach was based on the key assumption that the copolymer sequence can be described as a first-order Markov chain (Eq.(12) in Ref.\cite{kruger1987}).  By using this assumption, Kruger et.al. succeeded in reducing the original kinetic equations into  closed steady-state equations.
However, the validity of this assumption has not been proven rigorously or verified numerically. Moreover, how to generalize Kruger's approach to higher-order terminal models was unclear. So far as we know, the only attempt to extend Kruger logic to penultimate models with depropagation has been made by Li et al.\cite{hutchinson2005, hutchinson2006}.
In their works, however, the first-order Markov chain assumption which is valid only for terminal models, was inappropriately employed. This makes their penultimate model mathematically self-inconsistent (detailed discussion will be given in Section \ref{penultimate}). By far there are no well-established penultimate models with depropagation available in the literature.

Recently, the study on steady-state copolymerization also attracted attention from physicists who were interested to visualize the nonequilibrium copolymerization as information-generating process.  In Ref.\cite{gaspard2008,broeck2010,qianhong2009}, the zero-order copolymerization model with depropagation (named as Bernoullian model in this article, see Section \ref{bernoullian}) was introduced, without giving the derivation of the steady-state equations, to discuss some interesting issues (e.g., fidelity of DNA replication). In Ref.\cite{gaspard2014}, the first-order terminal model was discussed, similar to Kruger, under the assumption of first-order Markov chain. These works also put an emphasis on the thermodynamics of steady-state copolymerization and gave very general and interesting relations between the copolymer sequence entropy and the thermodynamic entropy production of the copolymerization system.
However, there still lacks of a systematic investigation on the steady-state kinetics and thermodynamics of any-order terminal model with depropagation.

In this article, we will generalize Kruger's Markov-chain assumption of the copolymer sequence distribution and suggest an unified mathematical approach to solve the steady-state kinetic equations of any-order terminal model with depropagation.  Based on the solution, we will also present a detailed discussion on the steady-state thermodynamics of copolymerization.

\section{Basic theory of steady-state copolymerization kinetics}
\subsection{Bernoullian model} \label{bernoullian}
As the simplest case of copolymerization, Bernoullian model (i.e., zero-order terminal model) assumes that the propagation and depropagation of monomers are independent of the terminal monomer unit. Although it's not a good model for real copolymerization systems, it can serve as a starting point of our discussion.
Below we investigate a two-component (A, B) system. Generalization to more complex cases (e.g., multi-component systems) will be given in later sections.

Denoting the propagation rate constants as $k^0_{\rm \tiny A}$ or $k^0_{\rm \tiny B}$, and depropagation rate constants as $\bar{k}_{\rm \tiny A}$ or $\bar{k}_{\rm \tiny B}$, we have the reaction scheme below
\begin{eqnarray}
{\sim}\cdot + {\rm A} \raisebox{-1.5ex}{$\stackrel{\stackrel{k^0_{\rm \tiny A}}{\textrm{\Large
$\rightleftharpoons$}}}{\textrm{\tiny $\bar{k}_{\rm \tiny A}$}}$}{\sim \rm A}\cdot,  \
{\sim}\cdot + {\rm B} \raisebox{-1.5ex}{$\stackrel{\stackrel{k^0_{\rm \tiny B}}{\textrm{\Large
$\rightleftharpoons$}}}{\textrm{\tiny $\bar{k}_{\rm \tiny B}$}}$}{\sim \rm B}\cdot  \nonumber
\end{eqnarray}

Imaging a single growing copolymer. $\sim \cdot$ represents the reactive end (i.e., the growing end, being either A$\cdot$ or B$\cdot$), and the occurrence probability of A$\cdot$ or B$\cdot$ at the terminal is denoted as $P_{\rm \tiny A}$ and $P_{\rm \tiny B}$ respectively.  We define $k_{\rm \tiny A} \equiv k^0_{\rm \tiny A}[{\rm A}]$, $k_{\rm \tiny B} \equiv k^0_{\rm \tiny B}[{\rm B}]$, [A], [B] are monomer concentrations in the environment which are constants during steady-state copolymerization. Supposing at some moment the copolymer contains $N_{\rm \tiny A}$ monomer A and $N_{\rm \tiny B}$ monomer B, the total number of monomers $N=N_{\rm \tiny A} + N_{\rm \tiny B}$. They all increase with time during copolymerization, and the corresponding kinetic equations are
\begin{eqnarray}
\dot{N}_{\rm \tiny A} &\equiv& J_{\rm \tiny A} = k_{\rm \tiny A} - \bar{k}_{\rm \tiny A} P_{\rm \tiny A}\nonumber\\
\dot{N}_{\rm \tiny B} &\equiv& J_{\rm \tiny B} = k_{\rm \tiny B} - \bar{k}_{\rm \tiny B} P_{\rm \tiny B}\label{NaNb}\\
\dot{N} &\equiv& J_{\rm tot} = J_{\rm \tiny A} + J_{\rm \tiny B}\nonumber
\end{eqnarray}
$J_{\rm \tiny A}$, $J_{\rm \tiny B}$ are respectively the overall incorporation rates of A and B. In steady-state copolymerization, ${\rm d} (N_{\rm \tiny A}/N)/{\rm d}t={\rm d}(N_{\rm \tiny B}/N)/{\rm d}t=0 $, or equivalently, $N_{\rm \tiny A}/N=\dot{N}_{\rm \tiny A}/\dot{N} = J_{\rm \tiny A}/J_{\rm tot}$ and $N_{\rm \tiny B}/N=\dot{N}_{\rm \tiny B}/\dot{N} = J_{\rm \tiny B}/J_{\rm tot}$.
So the overall occurrence probability of A or B in the copolymer can be expressed as $Q_{\rm A} \equiv N_{\rm \tiny A}/N = J_{\rm \tiny A} / J_{\rm tot}, Q_{\rm B} \equiv N_{\rm \tiny B}/N = J_{\rm \tiny B} / J_{\rm tot}$.

Higher order of chain-end sequence distribution $P_{i_n\cdots i_1}$( $i_m=$ A or B, $m=1,2,\cdots,n$. $i_n\cdots i_1$ denotes the chain-end sequence, with $i_1$ representing the terminal unit), and the total number of sequence $i_n\cdots i_1$ occurring in the copolymer chain $N_{i_n\cdots i_1}$ can be similarly defined. $\dot{N}_{i_n\cdots i_1} \equiv J_{i_n\cdots i_1} = k_{i_1}P_{i_n\cdots i_2} - \bar{k}_{i_1} P_{i_n\cdots i_1}$. In general, we have $P_{i_n\cdots i_1} = P_{{\rm A}i_n\cdots i_1} + P_{{\rm B}i_n\cdots i_1}$, $J_{i_n\cdots i_1} = J_{{\rm A}i_n\cdots i_1} + J_{{\rm B}i_n\cdots i_1}$. We also define $\widetilde{J}_{i_n\cdots i_1*} \equiv J_{i_n\cdots i_1{\rm A}} + J_{i_n\cdots i_1{\rm B}}$ and the overall sequence distribution $Q_{i_n\cdots i_1} \equiv N_{i_n\cdots i_1}/ N = J_{i_n\cdots i_1}/ J_{\rm tot} $.

The kinetic equations of $P_{i_n\cdots i_1}$ ($n\geq 1$) can be written as
\begin{eqnarray}
\dot{P}_{i_n\cdots i_1} = J_{i_n\cdots i_1} - \widetilde{J}_{i_n\cdots i_1*}  \label{Pini1-0}
\end{eqnarray}
For example,
\begin{eqnarray}
\dot{P_{\rm \tiny A}} &=& J_{\rm \tiny A} - \widetilde{J}_{{\rm A}*} = J_{\rm BA} - J_{\rm AB} \nonumber \\
&=& k_{\rm \tiny A} P_{\rm \tiny B} - k_{\rm \tiny B} P_{\rm \tiny A} + \bar{k}_{\rm \tiny B} P_{\rm AB}- \bar{k}_{\rm \tiny A} P_{\rm BA}
\end{eqnarray}
The existence of depropagation rates $\bar{k}_{\rm \tiny A}, \bar{k}_{\rm \tiny B}$ makes these equations hierarchically coupled and hard to be solved. Fortunately, for steady-state copolymerization $\dot{P}_{i_n\cdots i_1} = 0$ (for any $n\geq 1$), we can use the following truncation method to solve these equations.

In Bernoullian model, the steady-state copolymerization kinetics is determined only by $P_{\rm \tiny A}, P_{\rm \tiny B}$. This means that the coupled equations are redundant and can be reduced to equations of the two basic variables $P_{\rm \tiny A}, P_{\rm \tiny B}$. This reduction can be achieved by the following zero-order factorization conjecture of the chain-end sequence distribution, $P_{i_n\cdots i_1}=\prod^{n}_{m=1}P_{i_m}$, which leads to
\begin{eqnarray}
J_{i_n\cdots i_1} = \left( \prod^{n}_{m=2}P_{i_m} \right) J_{i_1}, \ \
\widetilde{J}_{i_n\cdots i_1*} = \left( \prod^{n}_{m=1}P_{i_m} \right) J_{\rm tot} \label{Jini1-0}
\end{eqnarray}
From the steady-state kinetic equation $0=\dot{P}_{i_n\cdots i_1} =J_{i_n\cdots i_1} - \widetilde{J}_{i_n\cdots i_1*}$, we can get
\begin{eqnarray}
\frac{J_{i_1}}{P_{i_1}} = J_{\rm tot}
\end{eqnarray}

Therefore, each of the coupled equations is reduced to the same steady-state equation of $P_{\rm \tiny A}, P_{\rm \tiny B}$,
\begin{eqnarray}
\frac{J_{\rm \tiny A}}{P_{\rm \tiny A}}  = \frac{J_{\rm \tiny B}}{P_{\rm \tiny B}}\   \label{JkPk}
\end{eqnarray}
Combining the normalization condition $P_{\rm \tiny A} + P_{\rm \tiny B} = 1$, we now obtain a set of closed equations
which gives the exact solution of the original kinetic equations (these steady-state equations have been used without derivation in \cite{broeck2010, qianhong2009}).

\begin{figure}[t]
\centering
\includegraphics[width=12cm]{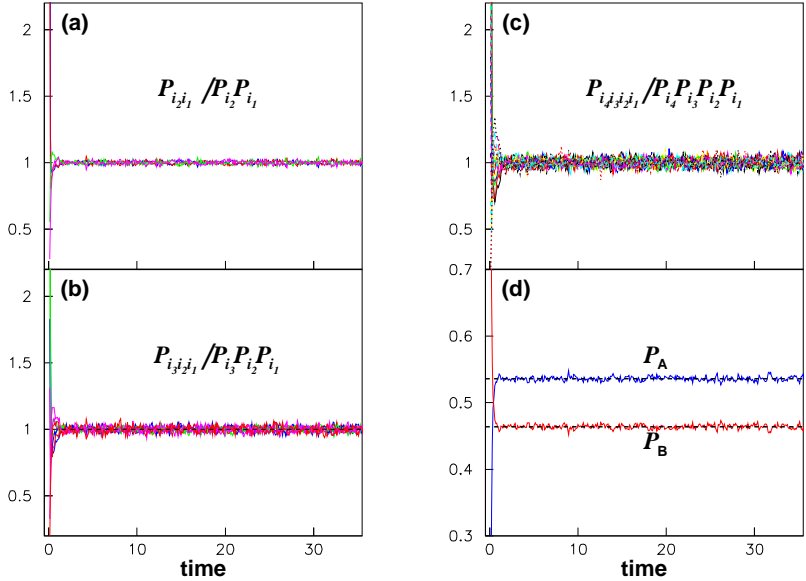}
\caption{Simulation verification of the zero-order factorization conjecture (a,b,c) and steady-state equation(d), with illustrative rate parameters $k_{\rm \tiny A}=4.0$, $k_{\rm \tiny B}=3.0$, $\bar{k}_{\rm \tiny A}=2.0$, $\bar{k}_{\rm \tiny B}=1.0$. The theoretical values of $P_{\rm A}, P_{\rm B}$ given by the steady-state equations are indicated as dash lines in (d). Here $i_m = {\rm A,B},\  m=1,2,3,4$. } \label{f1}
\end{figure}

Support of the factorization conjecture comes from the Monte-carlo simulations by using Gillespie algorithm \cite{gillespie1977,gillespie2007} (here the rate parameters are arbitrarily chosen). One can directly simulate the steady-state copolymerization from any given initial condition of $P_{\rm A}$, $P_{\rm B}$,  and obtain all the sequence statistics (e.g., the chain-end sequence distribution $P_{i_1}$, $P_{i_2i_1}$, \textit{etc.}) from a number of simulations.
For simplicity, we only check the factorizations $P_{i_2i_1}=P_{i_2}P_{i_1}$, $P_{i_3i_2i_1}=P_{i_3}P_{i_2}P_{i_1}$ and  $P_{i_4i_3i_2i_1}=P_{i_4}P_{i_3}P_{i_2}P_{i_1}$. As shown in Fig.\ref{f1}(a-c), for arbitrary choice of rate parameters $k_{\rm \tiny A}, k_{\rm \tiny B}, \bar{k}_{\rm \tiny A}, \bar{k}_{\rm \tiny B}$ (the only constraint on the parameters is that they should ensure $J_{\rm tot}>0$, i.e., the copolymer is growing), all the equalities hold when copolymerization reaches the unique steady state (which is determined only by rate parameters and independent of the choice of initial conditions). In Fig.\ref{f1}(d), we plot the time-evolution trajectories of $P_{\rm A}, P_{\rm B}$ given by the simulation and also indicate the steady-state values (shown by dash lines) of $P_{\rm A}, P_{\rm B}$ obtained by numerically solving Eq.(\ref{JkPk}), which also shows good agreement between the simulation and the theory.

\subsection{Terminal model}\label{terminal}
The so-called terminal model (i.e., the first-order terminal model), where the propagation and depropagation of monomer A and B are dependent on the identity of the terminal monomer unit, is a much more realistic model than Bernoullian model for real copolymerization systems. The reaction pathways in terminal model are
\begin{eqnarray}
&&\sim{\rm A}\cdot + {\rm A} \raisebox{-1.5ex}{$\stackrel{\stackrel{k_{\rm \tiny AA}}{\textrm{\Large
$\rightleftharpoons$}}}{\textrm{\tiny $\bar{k}_{\rm \tiny AA}$}}$}\sim{\rm AA}\cdot,\
\sim{\rm A}\cdot + {\rm B} \raisebox{-1.5ex}{$\stackrel{\stackrel{k_{\rm \tiny AB}}{\textrm{\Large
$\rightleftharpoons$}}}{\textrm{\tiny $\bar{k}_{\rm \tiny AB}$}}$}\sim{\rm AB}\cdot,\nonumber\\
&&\sim{\rm B}\cdot + {\rm A} \raisebox{-1.5ex}{$\stackrel{\stackrel{k_{\rm \tiny BA}}{\textrm{\Large
$\rightleftharpoons$}}}{\textrm{\tiny $\bar{k}_{\rm \tiny BA}$}}$}\sim{\rm BA}\cdot,\
\sim{\rm B}\cdot + {\rm B} \raisebox{-1.5ex}{$\stackrel{\stackrel{k_{\rm \tiny BB}}{\textrm{\Large
$\rightleftharpoons$}}}{\textrm{\tiny $\bar{k}_{\rm \tiny BB}$}}$}\sim{\rm BB}\cdot\nonumber
\end{eqnarray}

Defining
\begin{eqnarray}
J_{i_n\cdots i_2i_1} &\equiv& k_{i_2i_1}P_{i_n\cdots i_2} - \bar{k}_{i_2i_1}P_{i_n\cdots i_2i_1}\nonumber\\
\widetilde{J}_{i_n\cdots i_2i_1*} &\equiv& J_{i_n\cdots i_2i_1{\rm A}} + J_{i_n\cdots i_2i_1{\rm B}}\label{Jdefinition}
\end{eqnarray}
where $i_m = {\rm A, B} (m=1,2,\cdots ,n; \ n\geq 2)$,  we can write the corresponding kinetic equations for $P_{i_n\cdots i_2i_1}$ ($n\geq 1$) as below
\begin{eqnarray}
\dot{P}_{i_n\cdots i_2i_1}=J_{i_n\cdots i_2i_1}-\widetilde{J}_{i_n\cdots i_2i_1*} \label{Pini1-1}
\end{eqnarray}

The basic variables here are $P_{\rm AA}, P_{\rm AB}, P_{\rm BA}, P_{\rm BB}$, rather than $P_{\rm \tiny A}, P_{\rm \tiny B}$. Following the same logic in the previous section, we can reduce the hierarchically coupled equations Eq.(\ref{Pini1-1}) to an equivalent set of closed equations of $P_{\rm AA}, P_{\rm AB}, P_{\rm BA}, P_{\rm BB}$, by using the first-order factorization conjecture
\begin{eqnarray}
P_{i_n\cdots i_2i_1}= {\displaystyle\prod^{n}_{m=2}P_{i_mi_{m-1}}}\left[{\displaystyle\prod^{n}_{m=3}P_{i_{m-1}}}\right]^{-1}, \ n\ge3
\end{eqnarray}

Then the steady-state kinetic equations $\dot{P}_{i_n\cdots i_2i_1}=0$ ($n\geq 2$) are reduced to
\begin{eqnarray}
\frac{J_{i_2i_1}}{P_{i_2i_1}} = \frac{\widetilde{J}_{i_1*}}{P_{i_1}}
\end{eqnarray}
or equivalently,
\begin{eqnarray}
\frac{J_{{\rm A}i}}{P_{{\rm A}i}} = \frac{J_{{\rm B}i}}{P_{{\rm B}i}}  \label{JAkJk}
\end{eqnarray}
where $i={\rm A, B}$.

$\dot{P}_{\rm A} = J_{\rm \tiny A} - \widetilde{J}_{{\rm A}*}$ or $\dot{P}_{\rm B} = J_{\rm \tiny B} - \widetilde{J}_{{\rm B}*}$  (they are equivalent since $P_{\rm \tiny A} + P_{\rm \tiny B} =1$) leads to another steady-state equation $J_{\rm AB} = J_{\rm BA}$.  Finally we get four equations for four variables
\begin{eqnarray}
&&\frac{J_{\rm AA}}{P_{\rm AA}}= \frac{J_{\rm BA}}{P_{\rm BA}}, \ \
\frac{J_{\rm AB}}{P_{\rm AB}} = \frac{J_{\rm BB}}{P_{\rm BB}}, \ \
J_{\rm AB} = J_{\rm BA}, \nonumber\\
&&P_{\rm AA}+ P_{\rm AB}+ P_{\rm BA}+ P_{\rm BB}= 1\label{JakPak}
\end{eqnarray}
which gives the solution of the original kinetic equations.

\begin{figure}[t]
\centering
\includegraphics[width=12cm]{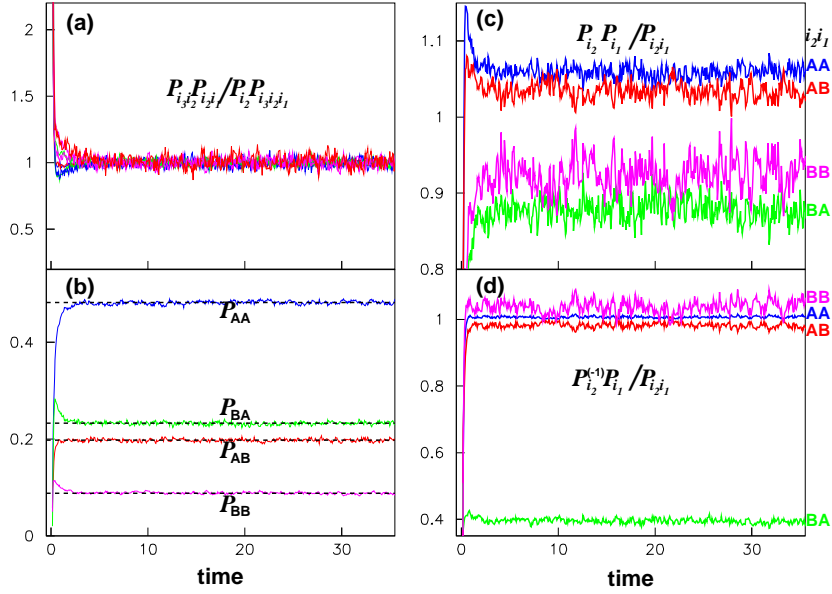}
\caption{Simulation verification of the first-order factorization conjecture (a) and steady-state equations (b), with illustrative rate parameters $k_{\rm AA}=6.0$, $k_{\rm AB}=2.0$, $k_{\rm BA}=4.0$, $k_{\rm BB}=1.0$, $\bar{k}_{\rm AA}=7.0$, $\bar{k}_{\rm AB}=5.0$, $\bar{k}_{\rm BA}=3.0$, $\bar{k}_{\rm BB}=1.0$. Theoretical values of $P_{\rm AA}$,$P_{\rm AB}$,$P_{\rm BA}$,$P_{\rm BB}$ given by steady-state equations are indicated as dash lines in (b). The zero-order factorization used in Bernoullian model fails in terminal model, as shown in (c). Direct factorization (see the text) also fails in terminal model, as shown in (d). Here $i_m = {\rm A,B},\  m=1,2,3$.   } \label{f2}
\end{figure}

The validity of the factorization $P_{i_3i_2i_1}=P_{i_3i_2}P_{i_2i_1}/P_{i_2}$ and the steady-state equations Eq.(\ref{JakPak}) can be checked by Monte-Carlo simulation. Fig.\ref{f2}(a) shows the factorization holds when copolymerization reaches steady state, and Fig.\ref{f2}(b) shows that the steady-state values of $P_{\rm AA}$, $P_{\rm AB}$, $P_{\rm BA}$, $P_{\rm BB}$ obtained by directly solving Eq.(\ref{JakPak}) are in good agreement with that given by the simulation (rate parameters used in the simulations are arbitrarily chosen as long as $J_{\rm tot}>0$ ).

It seems also possible in principle to use the zero-order factorization conjecture to reduce the original kinetic equations to closed steady-state equations. However, as shown by Fig.\ref{f2}(c), the zero-order factorization fails in terminal model, meaning that it's not applicable to terminal models. One may also suggest other factorization conjectures, for instance, the direct factorization $P_{ij}=P^{(-1)}_i P_j$ ($P^{(-1)}_i$ refers to the occurrence probability of monomer unit $i$ at the penultimate position. In fact, this conjecture does not result in closed steady-state equations). As indicated by Fig.\ref{f2}(d), this factorization also fails in terminal model, meaning that the correlation between the terminal unit and the penultimate unit can not be decoupled. In other words, one should take $P_{ij}$ as the basic variables to describe the terminal effect.

The first-order factorization conjecture $P_{i_3i_2i_1}=P_{i_3i_2}P_{i_2i_1}/P_{i_2}$ is actually equivalent to the first-order Markov-chain assumption used in Ref.\cite{kruger1987} and Ref.\cite{gaspard2014}. Defining transition probability $p(i_2| i_1) \equiv P_{i_2i_1}/P_{i_1}$, $p({\rm A}| i_1)+ p({\rm B}| i_1)=1 $, we now can rewrite the factorization conjecture as $P_{i_3i_2i_1}= p(i_3| i_2)p(i_2| i_1)P_{i_1}$.  It's worth noting that we have chosen $P_{i_2i_1}$, rather than $P_{i_1}$ and $p(i_2|i_1)$, as basic variables so as to represent the steady-state equations in a much simpler and more intuitive form.

\subsection{Penultimate model} \label{penultimate}
The reaction pathway of penultimate model (i.e., the second-order terminal model) can be expressed as
\begin{eqnarray}
\sim{\rm AA}\cdot + {\rm A} \raisebox{-1.5ex}{$\stackrel{\stackrel{k_{\rm \tiny AAA}}{\textrm{\Large
$\rightleftharpoons$}}}{\textrm{\tiny $\bar{k}_{\rm \tiny AAA}$}}$}\sim{\rm AAA}\cdot,&&
\sim{\rm AA}\cdot + {\rm B} \raisebox{-1.5ex}{$\stackrel{\stackrel{k_{\rm \tiny AAB}}{\textrm{\Large
$\rightleftharpoons$}}}{\textrm{\tiny $\bar{k}_{\rm \tiny AAB}$}}$}\sim{\rm AAB}\cdot,\nonumber\\
\sim{\rm AB}\cdot + {\rm A} \raisebox{-1.5ex}{$\stackrel{\stackrel{k_{\rm \tiny ABA}}{\textrm{\Large
$\rightleftharpoons$}}}{\textrm{\tiny $\bar{k}_{\rm \tiny ABA}$}}$}\sim{\rm ABA}\cdot,&&
\sim{\rm AB}\cdot + {\rm B} \raisebox{-1.5ex}{$\stackrel{\stackrel{k_{\rm \tiny ABB}}{\textrm{\Large
$\rightleftharpoons$}}}{\textrm{\tiny $\bar{k}_{\rm \tiny ABB}$}}$}\sim{\rm ABB}\cdot,\nonumber\\
\sim{\rm BA}\cdot + {\rm A} \raisebox{-1.5ex}{$\stackrel{\stackrel{k_{\rm \tiny BAA}}{\textrm{\Large
$\rightleftharpoons$}}}{\textrm{\tiny $\bar{k}_{\rm \tiny BAA}$}}$}\sim{\rm BAA}\cdot,&&
\sim{\rm BA}\cdot + {\rm B} \raisebox{-1.5ex}{$\stackrel{\stackrel{k_{\rm \tiny BAB}}{\textrm{\Large
$\rightleftharpoons$}}}{\textrm{\tiny $\bar{k}_{\rm \tiny BAB}$}}$}\sim{\rm BAB}\cdot,\nonumber\\
\sim{\rm BB}\cdot + {\rm A} \raisebox{-1.5ex}{$\stackrel{\stackrel{k_{\rm \tiny BBA}}{\textrm{\Large
$\rightleftharpoons$}}}{\textrm{\tiny $\bar{k}_{\rm \tiny BBA}$}}$}\sim{\rm BBA}\cdot,&&
\sim{\rm BB}\cdot + {\rm B} \raisebox{-1.5ex}{$\stackrel{\stackrel{k_{\rm \tiny BBB}}{\textrm{\Large
$\rightleftharpoons$}}}{\textrm{\tiny $\bar{k}_{\rm \tiny BBB}$}}$}\sim{\rm BBB}\cdot,\nonumber
\end{eqnarray}

Here the basic variables are $P_{i_3i_2i_1}$ ($i_3, i_2, i_1 = {\rm A, B}$). As in previous sections, we still define
\begin{eqnarray}
&&J_{i_n\cdots i_3i_2i_1} \equiv k_{i_3i_2i_1}P_{i_n\cdots i_3i_2} - \bar{k}_{i_3i_2i_1}P_{i_n\cdots i_3i_2i_1} \nonumber \\
&&\widetilde{J}_{i_n\cdots i_2i_1*} \equiv J_{i_n\cdots i_2i_1{\rm A}} + J_{i_n\cdots i_2i_1{\rm B}}
\end{eqnarray}
where $i_m = {\rm A, B} \ ( m=1,2,\cdots ,n; \ \ n\geq 3 )$.
The kinetic equation of $P_{i_n\cdots i_3i_2i_1}$ is
\begin{eqnarray}
\dot{P}_{i_n\cdots i_3i_2i_1}=J_{i_n\cdots i_3i_2i_1}-\widetilde{J}_{i_n\cdots i_3i_2i_1*} \label{Pini1-2}
\end{eqnarray}

To solve these equations, we take the following second-order factorization conjecture
\begin{eqnarray}
P_{i_n\cdots i_3i_2i_1}= {\displaystyle\prod^{n}_{m=3}P_{i_mi_{m-1}i_{m-2}}}\left[{\displaystyle\prod^{n}_{m=4}P_{i_{m-1}i_{m-2}}}\right]^{-1}, \ n\ge4
\end{eqnarray}
The steady-state kinetic equation $\dot{P}_{i_n\cdots i_3i_2i_1}=0$ ($n\geq 3$) can thus be reduced to
\begin{eqnarray}
\frac{J_{i_3 i_2 i_1}}{P_{i_3 i_2 i_1}} = \frac{\widetilde{J}_{i_2 i_1 *}}{P_{i_2 i_1}}  \label{JaimJim}
\end{eqnarray}
or equivalently
\begin{eqnarray}
\frac{J_{{\rm A} i_2 i_1}}{P_{{\rm A} i_2 i_1}} = \frac{J_{{\rm B} i_2 i_1}}{P_{{\rm B} i_2 i_1}}
\label{JaikPaik}
\end{eqnarray}
Now we have had five independent equations (Eq.(\ref{JaikPaik}), along with the normalization condition) for the eight variables. The rest three equations comes from the remaining kinetic equations of $P_{i_2i_1}$
\begin{eqnarray}
\dot{P}_{i_2i_1} = J_{i_2i_1} - \bar{J}_{i_2i_1 *} = 0
\end{eqnarray}
These four equations are not independent, due to the normalization condition $\sum P_{i_2i_1}=1$. So any three of them can be selected to form a closed set of equations of $P_{i_3i_2i_1}$, for instance,
\begin{eqnarray}
\frac{J_{\rm AAA}}{P_{\rm AAA}} &=& \frac{J_{\rm BAA}}{P_{\rm BAA}}, \ \frac{J_{\rm AAB}}{P_{\rm AAB}} = \frac{J_{\rm BAB}}{P_{\rm BAB}}, \nonumber\\
\frac{J_{\rm ABA}}{P_{\rm ABA}} &=& \frac{J_{\rm BBA}}{P_{\rm BBA}}, \ \frac{J_{\rm ABB}}{P_{\rm ABB}} = \frac{J_{\rm BBB}}{P_{\rm BBB}},\label{JaimPaim}\\
J_{\rm AA}&=&\widetilde{J}_{\rm AA*}, \ J_{\rm AB}=\widetilde{J}_{\rm AB*},  \nonumber\\
J_{\rm BB}&=&\widetilde{J}_{\rm BB*}, \ \sum_{i_3,i_2,i_1= \rm{A,B}} P_{i_3i_2i_1}=1\nonumber
\end{eqnarray}

\begin{figure}[t]
\centering
\includegraphics[width=12cm]{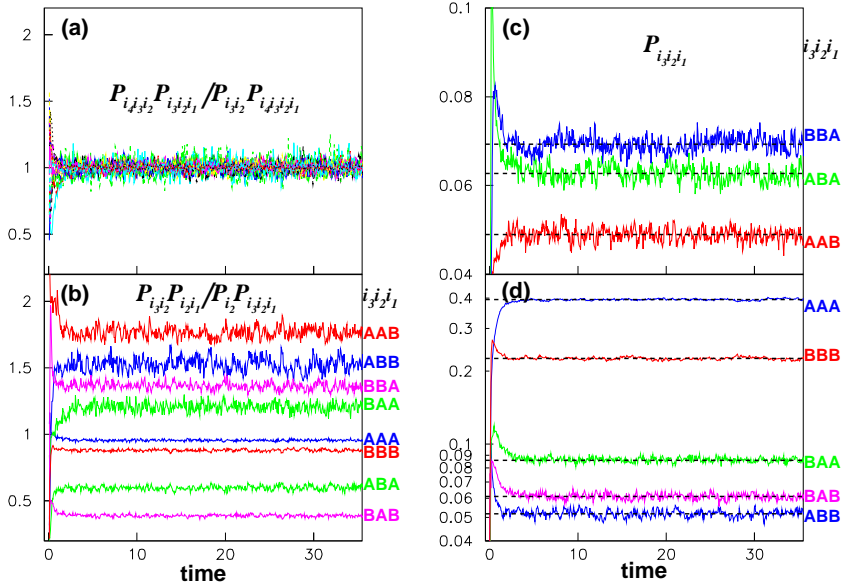}
\caption{Simulation verification of the second-order factorization conjecture (a) and steady-state equations (c,d), with illustrative rate parameters $k_{\rm AAA}=8.0$, $k_{\rm ABA}=4.0$, $k_{\rm BAA}=7.0$, $k_{\rm BBA}=2.0$, $k_{\rm AAB}=1.0$, $k_{\rm ABB}=3.0$, $k_{\rm BAB}=5.0$, $k_{\rm BBB}=6.0$, $\bar{k}_{\rm AAA}=7.0$, $\bar{k}_{\rm ABA}=3.0$, $\bar{k}_{\rm BAA}=8.0$, $\bar{k}_{\rm BBA}=4.0$, $\bar{k}_{\rm AAB}=5.0$, $\bar{k}_{\rm ABB}=1.0$, $\bar{k}_{\rm BAB}=6.0$, $\bar{k}_{\rm BBB}=2.0$. (b) implies that first-order factorization $P_{i_3i_2i_1}=P_{i_3i_2}P_{i_2i_1}/P_{i_2}$ is invalid in penultimate model. In (c,d), the theoretical values of $P_{i_3i_2i_1}$ given by steady-state equations are indicated by dash lines.} \label{f3}
\end{figure}

We checked the validity of $P_{i_4i_3i_2i_1}=P_{i_4i_3i_2}P_{i_3i_2i_1}/P_{i_3i_2}$ and the steady-state equations by Monte-Carlo simulations. Fig.\ref{f3}(a) shows the second-order factorization holds when copolymerization reaches the steady state,  Fig.\ref{f3}(c,d) shows a good agreement between the simulated values and the theoretical values of $P_{i_3i_2i_1}$.

It's worth noting that, as the zero-order factorization is inapplicable to first-order model, the first-order factorization is invalid for penultimate model, as indicated by Fig.\ref{f3}(b). In a recent paper, however, Li et al. invoked the first-order Markov chain assumption to solve their penultimate model\cite{hutchinson2005}.
According to our theory, Li's mathematical treatment is inappropriate. To account for penultimate effects, 8 variables $P_{i_3i_2i_1}\ (i_3,i_2,i_1={\rm A,B})$ are required. Li's model oversimplifies the problem to 6 variables (the two terminal probability $P_{i_1}$ and the four transition probability $p(i_2|i_1), i_2,i_1={\rm A,B}$, in terms of first-order Markov-chain assumption), and derived closed but overdetermined equations (Eq.(8-10) in Ref.\cite{hutchinson2005}) from the steady-state kinetic equations $\dot{P}_{i_2i_1}=0$. These equations are doubtful: two of the 16 rate parameters $k_{\rm AAA}, k_{\rm BBB}$ are totally absent from the original four kinetic equations $\dot{P}_{i_2i_1}= J_{i_2i_1}-\widetilde{J}_{i_2i_1*}$, meaning that these equations of 6-variables are inadequate to describe the penultimate effect.  Moreover, if Li's treatment is extended to higher-order kinetic equations (e.g., $\dot{P}_{i_3i_2i_1}=0$), self-inconsistency of their theory can be further uncovered. For instance, in the extreme case $\bar{k}_{i_3i_2i_1}=0$. Under the first-order Markov chain assumption, $\dot{P}_{\rm AAA}=k_{\rm AAA}P_{\rm BAA}-k_{\rm AAB}P_{\rm AAA}=0$ yields $k_{\rm AAA}P_{\rm BA}=k_{\rm AAB}P_{\rm AA}$, $\dot{P}_{\rm BAA}=k_{\rm BAA}P_{\rm BA}-(k_{\rm AAA}+k_{\rm AAB})P_{\rm BAA}=0$ yields $k_{\rm BAA}P_{\rm A}=(k_{\rm AAA}+k_{\rm AAB})P_{\rm AA}$. Combining these two equations will lead to the wrong conclusion $k_{\rm BAA}=k_{\rm AAA}$.
Therefore, the second-order but not first-order factorization conjecture is required for the penultimate model.

In general, if $m^{th}$-order factorization conjecture ($m<s$) is applied to the steady-state kinetic equations $\dot{P}_{i_n...i_1}=0$ ($n=1,2,...$) of $s^{th}$-order model (see the next section) , one can always obtain an overdetermined set of equations which is mathematically self-inconsistent.  On the other hand, higher-order ($m>s$) factorization conjecture is redundant for the $s^{th}$-order model.
We therefore conclude that the $s^{th}$-order model can only be appropriately described by $s^{th}$-order factorization conjecture.

\subsection{Higher-order terminal models} \label{higher}
The logic presented in previous sections can be directly generalized to higher-order terminal models.
Below we list the major results for $s^{th}$-order terminal model, i.e., the propagation and depropagation of $\rm{A/B}$ depend on the the last $s$ monomer units of the copolymer. Here the basic variables are $P_{i_{s+1} i_{s}\cdots i_1}$  ($2^{s+1}$ in total). We denote the propagation rates as $k_{i_{s+1} i_{s}\cdots i_1}$ and depropagation rates as $\bar{k}_{i_{s+1} i_{s}\cdots i_1}$, and also
\begin{eqnarray}
&&J_{i_n\cdots i_3i_2i_1} \equiv k_{i_{s+1}\cdots i_3i_2i_1}P_{i_n\cdots i_3i_2} - \bar{k}_{i_{s+1}\cdots i_3i_2i_1}P_{i_n\cdots i_3i_2i_1} \nonumber \\
&&\widetilde{J}_{i_n\cdots i_2i_1*} \equiv J_{i_n\cdots i_2i_1{\rm A}} + J_{i_n\cdots i_2i_1{\rm B}}
\end{eqnarray}
where $i_m = {\rm A, B}; \ \  m=1,2,\cdots ,n; \ \ n\geq s+1$.

The $s^{th}$-order factorization conjecture is
\begin{eqnarray}
P_{i_n\cdots i_1}= {\displaystyle\prod^{n}_{m=s+1}P_{i_mi_{m-1}\cdots i_{m-s}}}\left[{\displaystyle\prod^{n}_{m=s+2}P_{i_{m-1}\cdots i_{m-s}}}\right]^{-1}, \ n\ge s+2
\end{eqnarray}
The closed steady-state equations derived from $\dot{P}_{i_n\cdots i_1}= J_{i_n\cdots i_1} - \widetilde{J}_{i_n\cdots i_1 *}=0 \ (n\geq s+1)$ are the following $2^s$ equations
\begin{eqnarray}
\frac{J_{{\rm A} i_si_{s-1}\cdots i_1}}{P_{{\rm A} i_si_{s-1}\cdots i_1}} = \frac{J_{{\rm B} i_si_{s-1}\cdots i_1}}{P_{{\rm B} i_si_{s-1}\cdots i_1}} \end{eqnarray}
or equivalently,
\begin{eqnarray}
\frac{J_{i_{s+1}i_s\cdots i_1}}{P_{i_{s+1}i_s\cdots i_1}} = \frac{J_{i_si_{s-1}\cdots i_1}}{P_{i_si_{s-1}\cdots i_1}}
\label{sorderJP}
\end{eqnarray}

The kinetic equations $\dot{P}_{i_s\cdots i_1}= J_{i_s\cdots i_1} - \widetilde{J}_{i_s\cdots i_1 *}=0$ give other $2^s$ steady-state equations, from which any $2^s-1$ equations can be chosen. Combining the normalization condition $ \sum P_{i_{s+1} i_{s}\cdots i_1} = 1 $, we finally obtain a closed set of $2^{s+1}$ equations for $2^{s+1}$ variables.

The $s^{th}$-order factorization conjecture can be rewritten equivalently as $s^{th}$-order Markov chain, by defining the transition probability $p(i_{s+1}|i_s\cdots i_1) \equiv P_{i_{s+1}i_s\cdots i_1}/ P_{i_s\cdots i_1}$, $p( {\rm A}|i_s\cdots i_1)+ p({\rm B}|i_s\cdots i_1) = 1$.
Noting that $\widetilde{J}_{i_si_{s-1}\cdots i_1*}=J_{i_si_{s-1}\cdots i_1}$, the steady-state equations Eq.(\ref{sorderJP}) can be transformed into
\begin{eqnarray}
\frac{J_{i_{s+1} i_s\cdots i_1}}{J_{i_s\cdots i_1}}=\frac{P_{i_{s+1} i_s\cdots i_1}}{P_{i_s\cdots i_1}}
\end{eqnarray}
Since the overall sequence distribution $Q_{i_{s+1} i_s\cdots i_1}/Q_{i_s\cdots i_1}= J_{i_{s+1} i_s\cdots i_1}/J_{i_s\cdots i_1}$, the steady-state equations can be rewritten as
\begin{eqnarray}
\frac{Q_{i_{s+1} i_s\cdots i_1}}{Q_{i_s\cdots i_1}}=\frac{P_{i_{s+1} i_s\cdots i_1}}{P_{i_s\cdots i_1}}= p(i_{s+1}|i_s\cdots i_1)   \label{QP}
\end{eqnarray}
This simply means that the overall sequence distribution and chain-end sequence distribution can be described by the same $s^{th}$-order Markov chain.

It's also worth noting that the $s^{th}$-order model can reproduce $(s-1)^{th}$-order model if assuming $k_{{\rm A} i_s...i_1}=k_{{\rm B} i_s...i_1}= k_{i_s...i_1} $ and $\bar{k}_{{\rm A} i_s...i_1}= \bar{k}_{{\rm B} i_s...i_1} = \bar{k}_{i_s...i_1}$. By the $s^{th}$-order model, we have
\begin{eqnarray}
\frac{P_{{\rm A}i_s...i_2 i_1}}{P_{{\rm B}i_s...i_2 i_1}} = \frac{J_{{\rm A}i_s...i_2 i_1}}{J_{{\rm B}i_s...i_2 i_1}} = \frac{k_{i_s...i_2 i_1} P_{{\rm A}i_s...i_2} - \bar{k}_{i_s...i_2 i_1} P_{{\rm A}i_s...i_2i_1}}{k_{i_s...i_2 i_1} P_{{\rm B}i_s...i_2} - \bar{k}_{i_s...i_2 i_1} P_{{\rm B}i_s...i_2i_1}}
\end{eqnarray}
which yields
\begin{eqnarray}
\frac{P_{{\rm A}i_s...i_2 i_1}}{P_{{\rm B}i_s...i_2 i_1}} = \frac{P_{{\rm A}i_s...i_2}}{P_{{\rm B}i_s...i_2}}
\label{psps}
\end{eqnarray}
or equivalently
\begin{eqnarray}
\frac{P_{{\rm A}i_s...i_2 i_1}}{P_{{\rm A}i_s...i_2}}=\frac{P_{{\rm B}i_s...i_2 i_1}}{P_{{\rm B}i_s...i_2}} = \frac{P_{i_s...i_2 i_1}}{P_{i_s...i_2}}
\end{eqnarray}
This means $P_{i_{s+1}i_s...i_2 i_1} = P_{i_{s+1}...i_2}P_{i_s...i_1}/P_{i_s...i_2}$ which is exactly the  $(s-1)^{th}$-order factorization conjecture.

Eq.(\ref{psps}) also leads to
\begin{eqnarray}
\frac{P_{{\rm A}i_s...i_2}}{P_{{\rm B}i_s...i_2}}=\frac{J_{{\rm A}i_s...i_2 {\rm A}}}{J_{{\rm B}i_s...i_2 {\rm A}}} = \frac{J_{{\rm A}i_s...i_2 {\rm B}}}{J_{{\rm B}i_s...i_2 {\rm B}}}=\frac{\tilde{J}_{{\rm A}i_s...i_2*}}{\tilde{J}_{{\rm B}i_s...i_2*}}=\frac{J_{{\rm A}i_s...i_2}}{J_{{\rm B}i_s...i_2}}
\end{eqnarray}
which is exactly the steady-state equations of $(s-1)^{th}$-order model.

\section{Generalization to multi-component systems} \label{multicomp}
In real copolymerization systems, there might be multiple species of monomers. Generalization of the above kinetic theory to multi-component system is direct. Suppose there are $l$ species of monomer in the system (${\rm M}_1, {\rm M}_2,\cdots ,{\rm M}_l$).  In Bernoullian model, for instance, the basic variables are $P_{{\rm M}_j}\ (j=1,2,\cdots ,l)$. We have the following $l$ equations for the $l$ variables.
\begin{eqnarray}
\frac{J_{{\rm M}_1}}{ P_{{\rm M}_1}} = \frac{J_{{\rm M}_2}}{ P_{{\rm M}_2}}= \cdot\cdot\cdot = \frac{J_{{\rm M}_i}}{ P_{{\rm M}_i}}=\cdot\cdot\cdot= \frac{J_{{\rm M}_l}}{P_{{\rm M}_l}}, \ \
\sum^l_{i=1}P_{{\rm M}_i} = 1
\end{eqnarray}
where $J_{{\rm M}_i} \equiv k_{{\rm M}_i}- \bar{k}_{{\rm M}_i} P_{{\rm M}_i}$.

Generalization of higher-order models to multi-component system is similar(details not given here).

\section{Generalization to multi-step process} \label{multistep}
Different from cases discussed above where the propagation and depropagation are regarded as single-step process (e.g., in free radical copolymerization \cite{penlidis2005}),  bio-copolymerization such as DNA replication are often multi-step processes (e.g., \cite{qianhong2009}). In the latter case, the factorization conjectures can also be applied and similar steady-state equations can be derived. For simplicity and without loss of generality, we only discuss a simple case in Bernoullian model, as below
\begin{eqnarray}
&&{\sim}\cdot + {\rm A} \raisebox{-1.5ex}{$\stackrel{\stackrel{k_{\rm \tiny A1}}{\textrm{\Large
$\rightleftharpoons$}}}{\textrm{\tiny $\bar{k}_{\rm \tiny A1}$}}$}{\sim \rm A}^* \raisebox{-1.5ex}{$\stackrel{\stackrel{k_{\rm \tiny A2}}{\textrm{\Large
$\rightleftharpoons$}}}{\textrm{\tiny $\bar{k}_{\rm \tiny A2}$}}$}{\sim \rm A}\cdot,\nonumber\\
&&{\sim}\cdot + {\rm B} \raisebox{-1.5ex}{$\stackrel{\stackrel{k_{\rm \tiny B1}}{\textrm{\Large
$\rightleftharpoons$}}}{\textrm{\tiny $\bar{k}_{\rm \tiny B1}$}}$}{\sim \rm B}^* \raisebox{-1.5ex}{$\stackrel{\stackrel{k_{\rm \tiny B2}}{\textrm{\Large
$\rightleftharpoons$}}}{\textrm{\tiny $\bar{k}_{\rm \tiny B2}$}}$}{\sim \rm B}\cdot,\nonumber
\end{eqnarray}
here EA$^*$, EB$^*$ represent the intermediate states.  Now there are four possible states A, B, A$^*$, B$^*$ at the terminal, the corresponding probabilities are $P_{\rm A}, P_{{\rm A}^*}, P_{\rm B}, P_{{\rm B}^*}$. We now have kinetic equations very similar to Eq.(\ref{Pini1-0}), for instance,
\begin{eqnarray}
&&\dot{P}_m = J_{m^*} - \widetilde{J}_{m*} , \ \  i,m=A,B \nonumber \\
&&\dot{P}_{m^*} = J_{m^*} - J_{m}  \nonumber\\
&&J_{im^*} \equiv k_{m1}P_i-\bar{k}_{m1}P_{im^*}  \nonumber \\
&&J_{m^*} \equiv J_{{\rm A} m^*} + J_{{\rm B} m^*}    \\
&&\widetilde{J}_{m*} \equiv J_{m {\rm A}^*} + J_{m {\rm B}^*}   \nonumber \\
&&J_m \equiv k_{m2}P_{m^*}-\bar{k}_{m2}P_{m}  \nonumber
\end{eqnarray}

In steady state, $\dot{P}_{{\rm A}^*}=0$, $\dot{P}_{{\rm B}^*}=0$, i.e., we get two steady-state equations $J_{{\rm A}^*}=J_{\rm A}$, $J_{{\rm B}^*}=J_{\rm B}$. Furthermore, assuming $P_{ij}=\pi_{i}P_j$ and $P_{ij^*}=\pi_{i}P_{j^*}$, where $\pi_{i}\equiv P_i / (P_{\rm A} + P_{\rm B})$, one can again derive $J_{\rm A}/J_{\rm B} =\pi_{\rm A}/\pi_{\rm B}$, or equivalently $J_{\rm A}/P_{\rm A}=J_{\rm B}/P_{\rm B}$, from $\dot{P}_{\rm A}=0$ or $\dot{P}_{\rm B}=0$. Combining the normalization condition $P_{\rm A}+P_{{\rm A}^*}+P_{\rm B}+P_{{\rm B}^*}=1$, we thus have four equations for $P_{\rm A},P_{{\rm A}^*},P_{\rm B},P_{{\rm B}^*}$. It's worth noting that the above zero-order factorization conjectures generally hold for any multi-step processes in Bernoullian model.

Similarly, for the terminal model of multi-step processes, one can still assume the first-order factorization conjectures $P_{ijk}= P_{ij}P_{jk}/P_{j}$ and $P_{ijk^*}= P_{ij}P_{jk^*}/P_{j}$, etc.  This logic can be directly extended to any higher-order models.

\section{Steady-state thermodynamics} \label{thermo}
Recently, the steady-state thermodynamics of copolymerization has been discussed from the perspective of information theory \cite{gaspard2008,broeck2010,gaspard2014}. In these works, the authors proposed very general relations between the information of the copolymer sequence and the entropy production of the copolymerization system (e.g., Eq.(15) in Ref.\cite{gaspard2008}). Here we give explicit examples of such thermodynamic relations, based on the kinetic theory presented in previous sections. It should be pointed out first that the thermodynamics can only be well defined for special cases where propagation and depropagation are microscopically reversible (i.e., they proceed along the same reaction pathway),  whereas the kinetic theory is generally applicable even to cases in which propagation and depropagation proceed in different reaction pathways (e.g., in DNA replication, propagation is catalyzed by the synthesis domain of DNA polymerase, and depropagation is catalyzed by the editing domain of DNA polymerase). Below we assume that propagation and depropagation are microscopically reversible.

We start from the Bernoullian model, given that ${\rm A,B}$ are of identical concentration, i.e., $[{\rm A}]=[{\rm B}]=[{\rm M}]$. Details of higher-order models are given in Appendix.

In Bernoullian model, the averaged free energy dissipation per incorporation can be expressed as ( $RT$ is omitted for simplicity)
\begin{eqnarray}
\Delta G &=& \frac{\dot{S}}{J_{\rm tot}} \nonumber\\
\dot{S} &\equiv&
J_{\rm \tiny A}\textrm{ln}\left(\frac{k^0_{\rm \tiny A}[{\rm M}]}{\bar{k}_{\rm \tiny A}P_{\rm \tiny A}}\right)
+J_{\rm \tiny B}\textrm{ln}\left(\frac{k^0_{\rm \tiny B}[{\rm M}]}{\bar{k}_{\rm \tiny B}P_{\rm \tiny B}}\right) \\
J_{\rm \tiny A} &\equiv& k^0_{\rm \tiny A}[{\rm M}]- \bar{k}_{\rm \tiny A}P_{\rm \tiny A}, \ J_{\rm \tiny A} \equiv k^0_{\rm \tiny B}[{\rm M}]- \bar{k}_{\rm \tiny B}P_{\rm \tiny B} \nonumber\\
J_{\rm tot}&\equiv& J_{\rm \tiny A} + J_{\rm \tiny B}\nonumber
\end{eqnarray}
$\dot{S}$ is the instantaneous entropy production of the reaction system. It is non-negative by definition.

The unique equilibrium state can be defined by $J_{\rm \tiny A}=k^0_{\rm \tiny A}[{\rm M}]^{\rm eq}- \bar{k}_{\rm \tiny A}P^{\rm eq}_{\rm A}=0$ and $J_{\rm \tiny B}=k^0_{\rm \tiny B}[{\rm M}]^{\rm eq}- \bar{k}_{\rm \tiny B}P^{\rm eq}_{\rm B}=0$, which yield $[{\rm M}]^{\rm eq}= (K_{\rm A} + K_{\rm B})^{-1}$, $P^{\rm eq}_{\rm A} = K_{\rm A} [{\rm M}]^{\rm eq}$ and $P^{\rm eq}_{\rm B} = K_{\rm B} [{\rm M}]^{\rm eq}$, here $K_{\rm A}\equiv k^0_{\rm \tiny A}/ \bar{k}_{\rm \tiny A}$ and $K_{\rm B}\equiv k^0_{\rm \tiny B}/ \bar{k}_{\rm \tiny B}$ are the equilibrium constants of the $\rm {A,B}$ reaction pathway respectively. To drive the copolymerization process out of equilibrium, one should have
$[{\rm M}] > [{\rm M}]^{\rm eq} $.

In steady-state copolymerization, $J_{\rm \tiny A} / J_{\rm tot} =P_{\rm \tiny A}$, $J_{\rm \tiny B} / J_{\rm tot} =P_{\rm \tiny B}$, so we have
\begin{eqnarray}
\Delta G &=& P_{\rm \tiny A} \cdot \textrm{ln} \left( \frac{P^{\rm eq}_{\rm A}[{\rm M}]}{P_{\rm \tiny A}[{\rm M}]^{\rm eq}} \right)
   + P_{\rm \tiny B} \cdot \textrm{ln} \left( \frac{P^{\rm eq}_{\rm B}[{\rm M}]}{P_{\rm \tiny B}[{\rm M}]^{\rm eq}} \right)  \end{eqnarray}
this leads to the following equality
\begin{eqnarray}
\Delta G &=& \Delta \Psi - \Delta I  \nonumber\\
\Delta \Psi &\equiv& \textrm{ln}\left(\frac{[{\rm M}]}{[{\rm M}]^{\rm eq}}\right) \label{gipsi}\\
\Delta I &\equiv& P_{\rm \tiny A} \textrm{ln}\left(\frac{P_{\rm \tiny A}}{P^{\rm eq}_{\rm A}}\right) + P_{\rm \tiny B} \textrm{ln}\left(\frac{P_{\rm \tiny B}}{P^{\rm eq}_{\rm B}}\right)\nonumber
\end{eqnarray}

$ \Delta I$ is in the form of mutual information \cite{kl1951}. It's in fact the sequence information generated in the copolymerization process.
The information of a sequence of length $N$ is usually defined as Shannon entropy  $I_N = - \sum_{S_N} Q(S_N) \textrm{ln} Q(S_N)$, here the summation runs on all possible sequence $S_N = i_N i_{N-1} \cdots i_1 \ (i_m= \rm A,B) $ of length $N$, $Q(S_N) $ is the overall occurrence probability of $S_N$ in the copolymer sequence.

For the Bernoullian model, $Q(S_N)=\prod^{N}_{n=1}P_{i_n}$. One can prove $I_N = -N \left( P_{\rm \tiny A} \textrm{ln} P_{\rm \tiny A} + P_{\rm \tiny B} \textrm{ln} P_{\rm \tiny B} \right) $.
Change of the sequence information from the equilibrium state to the steady state is defined as $\Delta I_N = \sum Q(S_N) \textrm{ln} \left[ Q(S_N) / Q^{\rm eq}(S_N) \right]$, namely,

\begin{eqnarray}
\Delta I_N = N \left( P_{\rm \tiny A} \textrm{ln}\frac{P_{\rm \tiny A}}{P^{\rm eq}_{\rm A}}  + P_{\rm \tiny B} \textrm{ln}\frac{P_{\rm \tiny B}}{P^{\rm eq}_{\rm B}} \right) = N \Delta I
\end{eqnarray}
or
\begin{eqnarray}
\Delta I = \lim_{N \rightarrow \infty} \frac{1}{N}\sum_{S_N} Q(S_N) \textrm{ln} \frac{Q(S_N)}{Q^{\rm eq}(S_N)}
\end{eqnarray}

Hence, $\Delta I$ can be understood as information gain per incorporation. $\Delta \Psi$ can be regarded as the driving force to maintain the steady-state condition (i.e.,  to maintain the constant and nonequilibrium monomer concentration). The equality $\Delta \Psi = \Delta G + \Delta I$, which also holds for higher-order terminal models (details can be found in Appendix), has an intuitive meaning that the overall driving force ($\Delta \Psi > 0$) is partitioned into two parts, one to keep the polymerization reaction out of equilibrium ($\Delta G > 0$), the other to generate sequence information ($\Delta I > 0$).

The above theory can be readily generalized to multi-component systems (details not given here).

For multi-step processes, one can similarly define the free energy dissipation $\Delta G$ per incorporation. In the two-step process discussed in Section \ref{multistep}, for instance, $\Delta G$ can be written as (using the same notations as in Section \ref{multistep})
\begin{eqnarray}
\Delta G &=& \dot{S}/J_{\rm tot}    \nonumber\\
 J_{\rm tot}&\equiv& J_{\rm A} + J_{\rm B}  \\
 \dot{S} &\equiv& J_{\rm A}\textrm{ln}\left[\frac{k_{\rm A1}k_{\rm A2}(P_{\rm A}+P_{\rm B})}{\bar{k}_{\rm A1}\bar{k}_{\rm A2}P_{\rm A}}\right]
+ J_{\rm B}\textrm{ln}\left[\frac{k_{\rm B1}k_{\rm B2}(P_{\rm A}+P_{\rm B})}{\bar{k}_{\rm B1}\bar{k}_{\rm B2}P_{\rm B}}\right]  \nonumber\\
 &=& J_{{\rm A}^*}\textrm{ln}\left[\frac{k_{\rm A1}(P_{\rm A}+P_{\rm B})}{\bar{k}_{\rm A1}P_{{\rm A}^*}}\right]
+ J_{\rm A}\textrm{ln}\left[\frac{k_{\rm A2}P_{{\rm A}^*}}{\bar{k}_{\rm A2}P_{\rm A}}\right]  \nonumber\\
 &+& J_{{\rm B}^*}\textrm{ln}\left[\frac{k_{\rm B1}(P_{\rm A}+P_{\rm B})}{\bar{k}_{\rm B1}P_{{\rm B}^*}}\right]
+ J_{\rm B}\textrm{ln}\left[\frac{k_{\rm B2}P_{{\rm B}^*}}{\bar{k}_{\rm B2}P_{\rm B}}\right]
\nonumber
\end{eqnarray}
It's obvious that $\dot{S}$ is the instantaneous entropy production of the whole reaction systems, which is non-negative by definition.
Noting that $J_{\rm tot}$ is not defined as $J_{\rm tot}= J_{{\rm A}^*} + J_{\rm A} + J_{{\rm B}^*} + J_{\rm B}$, since the latter is the total flux of all the involved reactions but not the flux of incorporation, we have
\begin{eqnarray}
\Delta G = \pi_{\rm A}\textrm{ln}\left[\frac{k_{\rm A1}k_{\rm A2}}{\bar{k}_{\rm A1}\bar{k}_{\rm A2}\pi_{\rm A}}\right]
+ \pi_{\rm B}\textrm{ln}\left[\frac{k_{\rm B1}k_{\rm B2}}{\bar{k}_{\rm B1}\bar{k}_{\rm B2}\pi_{\rm B}}\right]
\end{eqnarray}
One can similarly define the information gain $\Delta I$ and the driving force $\Delta \Psi$,
\begin{eqnarray}
\Delta I \equiv \pi_{\rm \tiny A} \textrm{ln}\left(\frac{\pi_{\rm \tiny A}}{\pi^{\rm eq}_{\rm A}}\right) + \pi_{\rm \tiny B} \textrm{ln}\left(\frac{\pi_{\rm \tiny B}}{\pi^{\rm eq}_{\rm B}}\right) \nonumber \\
\Delta \Psi \equiv \textrm{ln}\left(\frac{[{\rm M}]}{[{\rm M}]^{\rm eq}}\right)
\end{eqnarray}
and also have $\Delta \Psi = \Delta G + \Delta I$.

The above discussion is based on the existence of the uniquely defined equilibrium state since $[{\rm A}]=[{\rm B}]$.  If $[{\rm A}]\neq[{\rm B}]$, one can still define the equilibrium state by $J_{\rm \tiny A}=k^0_{\rm \tiny A}[{\rm A}]^{\rm eq}- \bar{k}_{\rm \tiny A}P^{\rm eq}_{\rm A}=0$ and $J_{\rm \tiny B}=k^0_{\rm \tiny B}[{\rm B}]^{\rm eq}- \bar{k}_{\rm \tiny B}P^{\rm eq}_{\rm B}=0$, but the equilibrium state is not unique. Instead, there are infinite number of equilibrium states which satisfy $J_{\rm \tiny A}= 0$ and $J_{\rm \tiny B}= 0$.  In such cases, one can arbitrarily choose $[{\rm A}]^{\rm eq}$, $[{\rm B}]^{\rm eq}$, and then determine the corresponding $P^{\rm eq}_{\rm A}$, $P^{\rm eq}_{\rm B}$ to define the equilibrium state. The thermodynamic equality still holds, only with some modifications. For instance, for single-step process, we have
\begin{eqnarray}
\Delta \Psi &=&\Delta G + \Delta I \nonumber\\
\Delta G &\equiv& \frac{J_{\rm \tiny A}}{J_{\rm tot}}\textrm{ln}\left(\frac{k^0_{\rm \tiny A}[{\rm A}]}{\bar{k}_{\rm \tiny A}P_{\rm \tiny A}}\right)
+\frac{J_{\rm \tiny B}}{J_{\rm tot}}\textrm{ln}\left(\frac{k^0_{\rm \tiny B}[{\rm B}]}{\bar{k}_{\rm \tiny B}P_{\rm \tiny B}}\right) \ > 0 \nonumber\\
\Delta I &\equiv& P_{\rm \tiny A} \textrm{ln}\left(\frac{P_{\rm \tiny A}}{P^{\rm eq}_{\rm A}}\right) + P_{\rm \tiny B} \textrm{ln}\left(\frac{P_{\rm \tiny B}}{P^{\rm eq}_{\rm B}}\right) \ > 0 \\
\Delta \Psi &\equiv& P_{\rm \tiny A}\textrm{ln}\left(\frac{[{\rm A}]}{[{\rm A}]^{\rm eq}}\right) + P_{\rm \tiny B}\textrm{ln}\left(\frac{[{\rm B}]}{[{\rm B}]^{\rm eq}}\right)\nonumber
\end{eqnarray}

Last but not least, the information interpretation of $\Delta I$ is totally based on the $s^{th}$-order factorization conjecture of the $s^{th}$-order terminal model. Any lower-order factorization conjecture are incompatible with such an information interpretation. This gives an extra support of the factorization conjectures we used in the kinetic theory.

\section{Summary}
In this article, we proposed a systematic approach, based on Markov chain assumptions of the copolymer sequence distribution, to solve the unclosed kinetic equations of any-order terminal models with depropagation of steady-state copolymerization. The Markov chain assumptions were directly validated by Monte-carlo simulations, and the original kinetic equations were then reduced to closed steady-state equations which give the exact solution of the original equations. The derived steady-state equations were presented in an unified and intuitive form (e.g., Eq.(\ref{sorderJP})) which provide convenient tools to fit or explain experimental data.
This approach was also successfully generalized to more complex cases, e.g., multi-component systems and multi-step processes.
Furthermore, based on the steady-state equations and Markov chain assumptions, we derived a general thermodynamic equality in which the Shannon entropy of the copolymer sequence is explicitly introduced as part of the free energy dissipation of the whole copolymerization system. This not only offers extra support to the validation of the Markov chain assumptions, but also provides new insights to understand the copolymerization process from the perspective of information theory.

\section*{Appendix}
For $s^{th}$-order terminal model ($s \geq 1$), we define the unique equilibrium state by
\begin{eqnarray}
 J_{i_{s+1}i_s\cdots  i_1} &=& k^0_{i_{s+1}i_s\cdots  i_1}[{\rm M}]^{\rm eq} P^{\rm eq}_{i_{s+1}i_s\cdots  i_2} \nonumber\\
  &-&\bar{k}_{i_{s+1}i_s\cdots  i_1} P^{\rm eq}_{i_{s+1}i_s\cdots  i_1}=0
\end{eqnarray}
where $i_m = {\rm A, B} \ (m=1,2,\cdots ,s+1)$.
$[{\rm M}]^{\rm eq}, P^{\rm eq}_{i_{s+1}i_s\cdots  i_1}$ can be directly solved from these equations. They are functions only of equilibrium constants.

So we have
\begin{eqnarray}
\frac{k^0_{i_{s+1}i_s\cdots  i_1}}{\bar{k}_{i_{s+1}i_s\cdots  i_1}} = \frac{P^{\rm eq}_{i_{s+1}i_s\cdots  i_1}}{[{\rm M}]^{\rm eq}P^{\rm eq}_{i_{s+1}i_s\cdots  i_2}}   \label{eqcondition}
\end{eqnarray}

The averaged free energy dissipation is defined
\begin{eqnarray}
\Delta G \equiv \frac{1}{J_{\rm tot}} \sum_{\stackrel{i_m={\rm A,B}}{m=1,\cdots ,s+1} } J_{i_{s+1}\cdots  i_1} \textrm{ln}\left( \frac{k^0_{i_{s+1}\cdots  i_1}[{\rm M}] P_{i_{s+1}\cdots  i_2}}{\bar{k}_{i_{s+1}\cdots  i_1} P_{i_{s+1}\cdots  i_1}} \right)
\end{eqnarray}

Substituting Eq.(\ref{eqcondition}) into the above equation, we have
\begin{eqnarray}
\Delta G &=& \Delta \Psi - \Delta I  \nonumber\\
\Delta \Psi &\equiv &\frac{1}{J_{\rm tot}} \sum J_{i_{s+1}\cdots  i_1} \textrm{ln}\left( \frac{[{\rm M}]}{[{\rm M}]^{\rm eq}} \right) = \textrm{ln}\left( \frac{[{\rm M}]}{[{\rm M}]^{\rm eq}} \right) \label{deltaI}\\
\Delta I &\equiv& \frac{1}{J_{\rm tot}} \sum J_{i_{s+1}\cdots  i_1} \textrm{ln}\left( \frac{ P_{i_{s+1}\cdots  i_1}/ P_{i_{s+1}\cdots  i_2}}{P^{\rm eq}_{i_{s+1}\cdots  i_1}/ P^{\rm eq}_{i_{s+1}\cdots  i_2}} \right)
\nonumber
\end{eqnarray}

$\Delta I$ is also the information gain per incorporation step
\begin{eqnarray}
\Delta I = \lim_{N \rightarrow \infty} \frac{1}{N}\sum_{i_N\cdots i_1}Q_{i_N\cdots i_1} \textrm{ln} \frac{Q_{i_N\cdots i_1}}{Q^{\rm eq}_{i_N\cdots i_1}}
\end{eqnarray}

To prove this, we transform Eq.(\ref{deltaI}) into
\begin{eqnarray}
\Delta I &=& \frac{1}{J_{\rm tot}} \sum J_{i_{s+1}\cdots  i_1} \textrm{ln}\left( \frac{P_{i_{s+1}\cdots  i_1}/ P_{i_s\cdots  i_1}}{P^{\rm eq}_{i_{s+1}\cdots  i_1}/ P^{\rm eq}_{i_s\cdots  i_1}} \right)\nonumber\\
  &+& \frac{1}{J_{\rm tot}} \sum J_{i_{s+1}\cdots  i_1} \textrm{ln}\left( \frac{P_{i_s\cdots  i_1}/ P_{i_{s+1}\cdots  i_2}}{P^{\rm eq}_{i_s\cdots  i_1}/ P^{\rm eq}_{i_{s+1}\cdots  i_2}} \right)   \label{information}
\end{eqnarray}

Noticing that in the second term
\begin{eqnarray}
&&\sum_{i_{s+1},\cdots,i_1} J_{i_{s+1}\cdots i_1} \textrm{ln}\left( \frac{P_{i_s\cdots  i_1}}{P_{i_{s+1}\cdots  i_2}} \right)\nonumber\\
 &=& \sum_{i_{s+1},\cdots, i_1} J_{i_{s+1}\cdots  i_1} \textrm{ln} P_{i_s\cdots  i_1}   -\sum_{i_{s+1},\cdots, i_1} J_{i_{s+1}\cdots  i_1} \textrm{ln} P_{i_{s+1}\cdots  i_2}  \nonumber \\
&=& \sum_{i_s,\cdots, i_1} J_{i_s\cdots  i_1} \textrm{ln} P_{i_s\cdots  i_1}  -\sum_{i_{s+1},\cdots, i_2} \widetilde{J}_{i_{s+1}\cdots  i_2*} \textrm{ln} P_{i_{s+1}\cdots  i_2}  \nonumber \\
&=& \sum_{i_s,\cdots, i_1} J_{i_s\cdots i_1} \textrm{ln} P_{i_s\cdots  i_1}   -\sum_{j_s,\cdots, j_1} \widetilde{J}_{j_s\cdots  j_1*} \textrm{ln} P_{j_s\cdots  j_1} \nonumber \\
&=& \sum_{i_s,\cdots, i_1} \left( J_{i_s\cdots  i_1} - \widetilde{J}_{i_s\cdots  i_1*} \right) \textrm{ln} P_{i_s\cdots  i_1}=0
\end{eqnarray}
In the third step, we have substituted $i_{m}$ by $j_{m-1}$ ($m=2,3\cdots, s+1$). In the last step, we have used the steady-state conditions $J_{i_s\cdots  i_1} = \widetilde{J}_{i_s\cdots  i_1*}$.

Similarly, one can show
\begin{eqnarray}
\sum_{i_{s+1},\cdots, i_1} J_{i_{s+1}\cdots  i_1} \textrm{ln}\left( \frac{P^{\rm eq}_{i_s\cdots  i_1}}{P^{\rm eq}_{i_{s+1}\cdots  i_2}} \right)=0
\end{eqnarray}

Therefore, the second summation term in Eq.(\ref{information}) is zero. So we have
\begin{eqnarray}
\Delta I &=& \frac{1}{J_{\rm tot}} \sum J_{i_{s+1}\cdots  i_1} \textrm{ln}\left( \frac{P_{i_{s+1}\cdots  i_1}/ P_{i_s\cdots  i_1}}{P^{\rm eq}_{i_{s+1}\cdots  i_1}/ P^{\rm eq}_{i_s\cdots  i_1}} \right)  \nonumber
\end{eqnarray}

We define the transition probability as $p(i_{s+1}|i_s\cdots i_1) \equiv P_{i_{s+1}\cdots  i_1}/ P_{i_s\cdots  i_1}$. Since $ Q_{i_{s+1}\cdots  i_1} = J_{i_{s+1}\cdots  i_1}/ J_{\rm tot}$ and $ Q_{i_{s+1}\cdots  i_1}/ Q_{i_s\cdots  i_1}= P_{i_{s+1}\cdots  i_1}/ P_{i_s\cdots  i_1}=p(i_{s+1}|i_s\cdots i_1) $ (Eq.(\ref{QP})), we rewrite the above equation as
\begin{eqnarray}
&&\Delta I=\sum_{i_{s+1},\cdots, i_1} Q_{i_{s+1}\cdots  i_1} \textrm{ln}\left( \frac{p(i_{s+1}|i_s\cdots i_1)}{p^{\rm eq}(i_{s+1}|i_s\cdots i_1)} \right)  \label{inforgain}\\
&=&\sum_{i_{s+1},\cdots, i_1} Q_{i_s\cdots  i_1} p(i_{s+1}|i_s\cdots i_1) \textrm{ln}
\left( \frac{p(i_{s+1}|i_s\cdots i_1)}{p^{\rm eq}(i_{s+1}|i_s\cdots i_1)} \right)  \nonumber \\
&=&\sum_{i_s,\cdots, i_1} Q_{i_s\cdots  i_1}
\left[ p({\rm A}|i_s\cdots i_1) \textrm{ln}\left( \frac{p({\rm A}|i_s\cdots i_1)}{p^{\rm eq}({\rm A}|i_s\cdots i_1)} \right) \right.  \nonumber \\
&& \left. + p({\rm B}|i_s\cdots i_1) \textrm{ln}\left( \frac{p({\rm B}|i_s\cdots i_1)}{p^{\rm eq}({\rm B}|i_s\cdots i_1)} \right)\right] \nonumber
\end{eqnarray}

Since $p({\rm A}|i_s\cdots i_1)+ p({\rm B}|i_s\cdots i_1) =1 $ and $p^{\rm eq}({\rm A}|i_s\cdots i_1) + p^{\rm eq}({\rm B}|i_s\cdots i_1) =1 $, it can be readily proven that $\Delta I$ is nonnegative. Furthermore, one can show $\Delta I$ is exactly the information gain per incorporation.
The information of a sequence of length $N$ is $\Delta I_N = \sum Q(S_N) \textrm{ln} \left( Q(S_N) / Q^{\rm eq}(S_N) \right)$.
Since $Q_{i_N\cdots i_1}= \left[\prod^{N}_{n=s+1}p(i_n|i_{n-1}\cdots i_{n-s}) \right] Q_{i_s\cdots  i_1}$, we have
\begin{eqnarray}
\Delta I_N &=& \sum_{i_N,\cdots, i_1} Q_{i_N\cdots i_1} \left[ \sum^{N}_{n=s+1}\textrm{ln} \frac{p(i_n|i_{n-1}\cdots i_{n-s})}{p^{\rm eq}(i_n|i_{n-1}\cdots i_{n-s})} \right]\nonumber\\
&+&\sum_{i_s,\cdots, i_1} Q_{i_s\cdots i_1} \textrm{ln} \left( \frac{Q_{i_s\cdots i_1}}{Q^{eq}_{i_s\cdots i_1}} \right)
\label{DeltaIN}
\end{eqnarray}

The first summation term can be rewritten as
\begin{eqnarray}
&& \sum^{N}_{n=s+1} \left[ \sum_{i_n,\cdots, i_{n-s}}  \left( \sum_{\stackrel{i_{N},\cdots, i_{n+1},}{i_{n-s-1},\cdots, i_1}} Q_{i_N\cdots i_1} \right)\textrm{ln} \frac{p(i_n|i_{n-1}\cdots i_{n-s})}{p^{\rm eq}(i_n|i_{n-1}\cdots i_{n-s})}  \right] \nonumber \\
&& = \sum^{N}_{n=s+1} \left[ \sum_{i_n,\cdots, i_{n-s}}  Q_{i_n\cdots i_{n-s}} \textrm{ln} \frac{p(i_n|i_{n-1}\cdots i_{n-s})}{p^{\rm eq}(i_n|i_{n-1}\cdots i_{n-s})}  \right] \nonumber \\
&& = (N-s) \sum_{i_{s+1},\cdots, i_1} Q_{i_{s+1}\cdots  i_1} \textrm{ln} \left(\frac{p(i_{s+1}|i_s\cdots i_1)}{p^{\rm eq}(i_{s+1}|i_s\cdots i_1)} \right)
\end{eqnarray}

The second term of Eq.(\ref{DeltaIN}) is comparatively negligible. Hence the information gain per incorporation is
\begin{eqnarray}
\Delta I &=& \lim_{N \rightarrow \infty} \frac{1}{N} \Delta I_N  \nonumber \\
&=& \sum_{i_{s+1},\cdots, i_1} Q_{i_{s+1}\cdots  i_1} \textrm{ln} \left(\frac{p(i_{s+1}|i_s\cdots i_1)}{p^{\rm eq}(i_{s+1}|i_s\cdots i_1)} \right)
\end{eqnarray}

This is exactly Eq.(\ref{inforgain}).

\section*{Acknowledgments}
The authors thank the financial support by the National Basic Research Program of China (973 program, No.2013CB932804) and National Natural Science Foundation of China (No.11105218, No.91027046).

\end{document}